**Title**: The gene function prediction challenge: large language models and knowledge graphs to the rescue

**Authors**: Rohan Shawn Sunil[1], Shan Chun Lim[1], Itharajula Manoj[1], Marek Mutwil[1*]

[1]School of Biological Sciences, Nanyang Technological University, 60 Nanyang Drive, Singapore, 637551, Singapore

*Corresponding author:
Marek Mutwil
School of Biological Sciences,
Nanyang Technological University, 60 Nanyang Drive,
637551, Singapore,
Singapore
Email: mutwil@ntu.edu.sg

**Abstract**

Elucidating gene function is one of the ultimate goals of plant science. Despite this, only ~15% of all genes in the model plant *Arabidopsis thaliana* have comprehensively experimentally verified functions. While bioinformatical gene function prediction approaches can guide biologists in their experimental efforts, neither the performance of the gene function prediction methods nor the number of experimental characterisation of genes has increased dramatically in recent years. In this review, we will discuss the status quo and the trajectory of gene function elucidation and outline the recent advances in gene function prediction approaches. We will then discuss how recent artificial intelligence advances in large language models and knowledge graphs can be leveraged to accelerate gene function predictions and keep us updated with scientific literature.

**Why is gene function prediction needed?**

Understanding the functions of plant genes is essential to address the challenges of global food security, sustainable agriculture, and environmental challenges. For example, basic research into gene function allowed the engineering of plants with higher pathogen resistance [1,2], decreased energy requirements for growth [3,4], and decreased fertilizer requirements [5,6]. However, we still need to learn about the function of most genes, which precludes us from engineering plants to address the sustainability challenge. There are ~500,000 species in the plant kingdom, but fewer than 300 have chromosome-scale genomes [7]. The diploid flowering *Arabidopsis thaliana* is the most studied plant, containing ~30,000 genes, but only 15% are comprehensively characterized (Figure 1). The gene knowledge gap is even more exaggerated in food crops, as the percentage of partially characterized genes precipitously drops to single digits: 2% for rice and 0.6% for maize [8].

We can partially address these challenges with computational gene function predictions to guide experimental approaches [9]. However, while invaluable, these approaches often do not provide accurate predictions, and their performance has not improved much in recent years [10], indicating the need for a paradigm change. Fortunately, in the last four years, we have seen an explosion of advances in Artificial Intelligence (AI) along with the affordable and large-scale data generation needed to train these models. AI models can accurately predict Gene Ontology terms [11], protein structures [12], protein-protein interactions [13], gene expression from DNA sequence [14], alternative splicing [15], and other aspects of gene function [16]. However, all of these approaches require a gold standard comprising experimentally verified genes, which is still far from exhaustive.

Here, we will cover the current status, biases, and trajectories of gene function and the various computational approaches used to predict gene function. We will then show how large language models and knowledge graphs can vastly boost gene function prediction and how researchers use existing knowledge.

**How much do we know about gene function in plants?**
To better understand the current state and trends in gene function studies, we conducted a meta-analysis of experimentally verified gene functions and the journals that report them.

*Progress and Trajectory of Gene Function Discovery*
The number of papers reporting gene functions peaked in 2008 but has since dropped dramatically to levels seen before the publication of the *Arabidopsis thaliana* genome

(Figure 1a). Surprisingly, the number of publications using '*Arabidopsis thaliana*' as a query on NCBI has plateaued since around 2013 and has even increased in recent years (Figure 1a, red line). This disparity between the reported gene functions and the number of published articles might be driven by changing funding priorities, shifting from fundamental science to application-oriented goals such as food security and biofuels [17]. This shift is supported by the observation that funding-intensive sequencing experiments focus more on crops and stress responses [18]. Despite the drop in reported gene functions, only ~12.7% (3691 out of 29,001) of Arabidopsis genes are fully characterized by all three domains of gene ontology (Figure 1b). However, the number of genes with all three GO domains has increased >2-fold since 2014 (from ~5% to ~12.7%) [9], indicating increased attention to partially characterized genes. In summary, studies reporting new gene functions have declined dramatically over the past 13 years, which has also been observed in bacteria and yeast [19,20].

*Bias in Gene Characterization*

The number of studies per gene shows a high bias. Most genes have only one PubMed ID reporting their function (~10,000 genes have only one paper), while a few genes have over 40 papers describing their functions (Figure 1c). The top ten most-characterized genes are involved in processes such as photomorphogenesis (*COP1, HY1, AT-PHH1*), the circadian clock (*CRY1*), development (*ASK1, KIN10, RGA*), hormonal signaling (*BAK1, ATOST1*), and other functions (*ATMAPK6*).

*Journals Reporting Gene Functions*

The majority of gene functions are reported in specialized plant journals such as Plant Physiology (established in 1926, impact factor 7.4, 31.8k Twitter followers), Plant Cell (1989, impact factor 10, 35.8k followers), and Plant Journal (1991, impact factor 6.2, 26.2k followers). The fourth and fifth most frequently used journals are generalist, non-plant-specific journals: PNAS (1914, impact factor 9.4, 182k followers) and Journal of Biological Chemistry (1905, impact factor 5.4, 21.8k followers). These journals are likely popular due to their long-standing reputation, relatively high impact factors, significant reach within the scientific community, and innovative approaches to publishing [21].

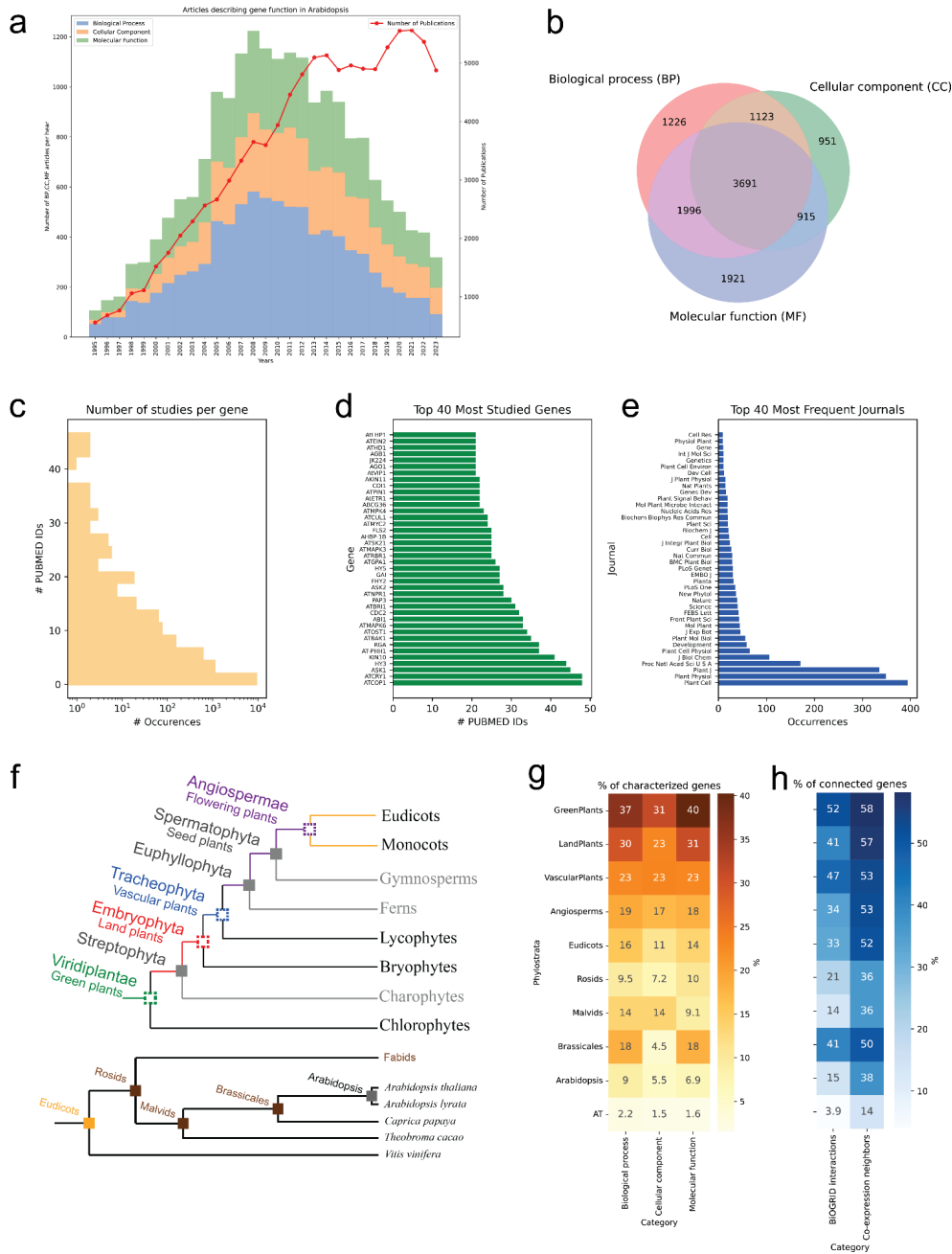

**Figure 1. Gene function information for *Arabidopsis thaliana*.** a) The x-axis shows the years, while the left y-axis indicates the number of papers reporting experimentally verified gene functions for Gene Ontology domains Molecular Function (MF, blue), Cellular Component (CC, orange) and Biological Process (BP, green). The right y-axis indicates the total number of publications found on NCBI for 'Arabidopsis thaliana'. b) The Venn diagram shows the overlap of experimentally verified functions for BP, CC and MF domains. c) The number of papers reporting new functional information for each gene. The y-axis indicates the number of papers per gene, while the x-axis represents the number of genes having a given number of papers. d) The number of papers (x-axis,

PubMed IDs) for the top 40 most studied genes. e) The number of journals (x-axis) reporting gene functions. f) The cladogram of plant clades. g) The percentage of experimentally characterized genes for BP, CC, MF (columns) and different phylostrata (rows). The phylostrata are sorted from oldest (GreenPlants) to youngest (AT: *Arabidopsis thaliana*). h) The percentage of *Arabidopsis thaliana* genes connected to experimentally characterized genes in BioGRID interaction (left column) and CoNekT (https://conekt.sbs.ntu.edu.sg/) database. The code to generate the figure is found at https://bit.ly/3YBh2JP.

*Age-specific Biases in Gene Characterization and Function Prediction*

We observed a negative association between gene age and functional characterization. Genes found in all land plants tend to be more characterized than genes found only in the Arabidopsis genus (Figure 1f-g) [22]. While it is unclear why this is so, we speculate that early plant research often began by identifying mutants with detectable phenotypes [23], and older genes tend to have more basal, essential functions [22], resulting in stronger phenotypes. In contrast, younger genes may function more specifically (organ/tissue/cell/condition-specific) [24] and exhibit milder phenotypes that are often missed in screens.

Gene function predictions operate on the guilt-by-association principle, using data such as gene co-expression and protein-protein interactions (PPI) to link functionally related genes [9]. As previously reported [25], older genes tend to be more connected to already characterized genes in PPI and co-expression networks, making it easier to predict their functions (Figure 1h). Conversely, newer genes show fewer connections to the characterized genes (Figure 1h), making it more difficult to predict and characterize their function.

**Current approaches to predict gene functions**

All gene function prediction approaches require three components: (i) omics data (genomes, proteomes, transcriptomes) used to identify similarities of gene features, (ii) a collection of experimentally verified genes (the gold standard, e.g., in the form of GO terms) that can serve as a source of functional knowledge and (iii) a computational method that uses the gold standard to functionally annotate genes with similar omic features (Figure 2) [9]. Consequently, all three components are essential for the quality of gene function predictions. Omics data, especially high-throughput sequencing techniques, have generated vast amounts of DNA and RNA sequences across hundreds of species [26,27].

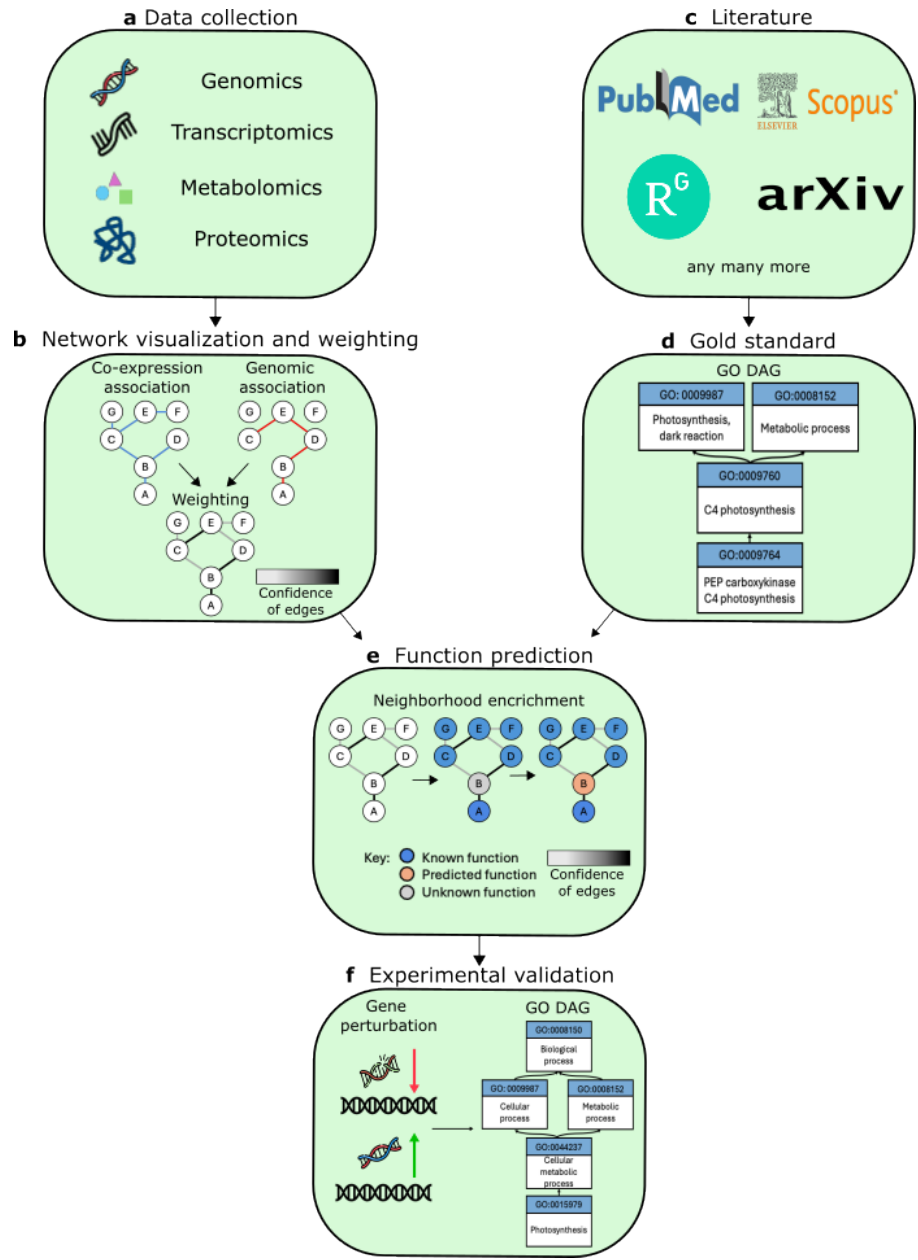

**Figure 2. Integrative workflow for gene function prediction using omics data and literature-validated standards.** The gene function prediction workflow involves several key steps: a) Data collection: Gather various omics data, including mRNA expression, protein-protein and genetic interactions, to obtain an overview of gene properties. b) Network visualization and weighting: Construct networks to visualize gene interactions through co-expression analysis, genomic context associations and sequence comparisons while assigning weights to these interactions to enhance confidence in the connections. c) Literature: Various research publication platforms provide access to valuable scientific discoveries. d) Gold standard: Experimentally validated genes found in

the literature are recognised as the gold standard. These genes are mapped to specific GO terms, which can be organized into a Directed Acyclic Graph (DAG) to represent and analyze gene relationships. e) Function Prediction: Use neighborhood enrichment or clustering methods to predict gene functions by leveraging integrated data and similarities to known genes. f) Experimental Validation: Test predictions experimentally, such as conducting gene knockouts to observe phenotypic changes.

Gene function prediction methods were initially sequence similarity centric (e.g., BLAST [28]), but in the last 15 years, have used advanced techniques like co-expression analysis, network-based methods, Machine Learning (ML), multi-omics integration, and deep learning [29]. Network-based methods use protein-protein interactions (PPI), co-expression networks and gene regulatory networks (GRNs) to predict gene functions [30]. Multi-omics approaches integrate genomics, transcriptomics, proteomics, and metabolomics to achieve synergistically better performance than methods based on a single type of data [31]. Machine Learning models, such as Random Forest (RF) and Support Vector Machine (SVM) [32], can utilize the multi-omics data to arrive at predictions but require hand-crafted biological features, which can be time-consuming to construct and are prone to human bias. To remedy this, deep learning approaches such as Convolutional Neural Networks (CNNs) [33] and Graph Convolutional Networks (GCNs) [34] were developed and can identify the relevant features from raw data. Recently, significant advances in large language models (LLMs), such as scGPT [35] (single cell Generative Pre-trained Transformer), have shown improvements in the precision of cell type annotation in humans, identifying functionally related genes, predicting responses to unseen genetic perturbations, and enabling multi-omic integration. AgroNT [36] has shown state-of-the-art predictions for regulatory annotations, promoter/terminator strength, and tissue-specific gene expression in plants. Additionally, interpretable DeepLearning models [37] have shown remarkable efficacy in predicting plant gene expression profiles, revealing cross-species regulatory features from DNA sequences.

While gene function prediction tools are invaluable in supporting experimental biologists, they still need to be improved. The Critical Assessment of Functional Annotation consortium (CAFA) revealed only a small improvement in gene function prediction methods in 2019 [10]. The challenges are that the pace of gene prediction outstrips functional characterisation [38], and predictions are biased towards well-understood pathways [39].

Thus, effective gene function prediction approaches will require high-quality multi-omics data, multimodal models able to integrate this data, and, paradoxically, a large corpus of characterized genes. While new data and methods will require resources to develop, the availability of the gold standard data can be vastly improved with current technology quickly due to advances in large language models and knowledge graphs.

**How can we use LLMs to expand our gold standard?**

*The bottleneck of gene function prediction*

The gold standard data is typically in tabulated Gene Ontology terms, making it suitable as input to gene function prediction methods. The generation of the gold standard relies on manual curation of the scientific literature, which is a labor-intensive process requiring a specialist that can bridge computational and experimental biology. Consequently, the size of the gold standard is primarily limited by the availability of human experts to read a research article and translate the findings in the paper (e.g., 'We used a pulldown assay to show that gene X and gene Y are interacting') to a corresponding gene ontology term ('gene X-> GO:0005515: protein binding', 'gene Y-> GO:0005515: protein binding', 'evidence code: Inferred from Physical Interaction (IPI)'). Thus, keeping up with novel discoveries can be slow and labor-intensive and likely result in a significant gap between the information available in journals and the gold standard for gene function predictions. Furthermore, current databases are typically tailored towards specific species (e.g. TAIR [40], MaizeGDB [41]) or specific biological properties (e.g. BioGRID for interactions [42], TAIR for gene ontologies [40]), each with their own set of schema, further complicating the utilization of the characterized gene functions.

*Use of large language models to build knowledge graphs*

Recently, automated forms of text-mining from literature are being used in the form of pre-trained LLMs, such as Bidirectional Encoder Representations from Transformers (BERT) and Generative Pre-trained Transformer (GPT) [43–45]. These sophisticated AI models are trained on large, diverse datasets to interpret and generate human-readable text. LLMs can be fine-tuned for task-specific datasets, such as BioGPT, BioBERT and SciBERT in scientific domains [46–48], or incorporated into tools and pipelines [49,50]. They can extract relationships and entities from unstructured text with minimal task-specific training data, significantly speeding up the data curation process. The output would be a knowledge graph (KG), which is structured data in the form of triples (consisting of a subject, predicate and object). Remarkably, LLMs can build knowledge graphs with a simple prompt, such as 'build a knowledge graph from this text', and identify the triples accurately [51].

The output is a directed graph ascribing semantic relationships (e.g., regulation, interaction, synthesis) between two connected entities (genes, metabolites, organs, perturbations), facilitating data integration and information retrieval. In plant biology, utilization of KGs is still an emerging trend, with varying levels of sophistication in their data curation methods. Stress Knowledge Map is a database manually curated by experts to showcase plant interactions and stress-specific responses [52]. AgroLD presents multi-omics data on plants collated from over 100 existing databases, integrated using Semantic Web technologies [53]. Meanwhile, PlantConnectome uses the natural language processing capabilities of GPT to mine and construct a KG spanning the breadth of results reported in ~100,000 plant literature abstracts [51]. The KG reported in PlantConnectome could uncover relationships previously unreported in databases such as BioGRID and AGRIS [42,54], demonstrating high accuracy (Figure 3a).

While KGs do not explicitly assign Gene Ontology terms, we propose that the edges that capture associations of genes to other entities (genes, metabolites, cellular compartments, biological processes) can tremendously simplify the manual assignment of GO terms or serve as an input to already established natural language processing methods that automatically assign terms from text [55].

**Figure 3. Knowledge graphs as tools to store information and predict gene functions.** a) Knowledge graph of CESA1, taken from PlantConnectome: http://plant.connectome.tools/alias/CESA1. Each node represents different entities, while the edges depict their relationships. Node and edge colors indicate the types of entities and relations. b) KGs can be used as input to generate KG embeddings using various models that can be translation-based, tensor factorisation-based or neural network-based. Proximity between vectors represents the strength of connection or similarity of characteristics between the entities. Potential relationships can be inferred by employing techniques that analyze proximity and directionality between entity vector embeddings.

**Knowledge graphs as predictive tools**

While KGs typically serve as knowledge bases for storing and retrieving information, it is possible to identify new relationships between entities by link prediction methods [56]. Link prediction methods use KG embeddings, where entities and relationships can be represented in a low-dimensional vector space whilst retaining the original topology of the graph (Figure 3b) [57]. Embeddings can be calculated with more intuitive methods, where the second entity's vector is the sum of the first entity's vector and the relationship vector [58]. In contrast, other methods represent the KG as a multi-dimensional adjacency matrix (tensor) that can be factorized and decomposed into low-dimensional vectors corresponding to its entities and relationships [59]. Alternatively, neural network-based methods use layers of computation that adjust parameters like weights and biases to capture and represent intricate graphical structures, including features of entities and relationships [56]. Two popular models are GCNs [60], which aggregate information from a node's neighbors to update its embedding, capturing detailed local information, and CNNs, which apply convolutional filters to extract local patterns and features [61]. These features are then distilled into compact vector embeddings representing the graph's hierarchical relationships.

Vector embeddings are mathematical representations of entities and their relationships. Intuitively, two entities with similar properties should have similar vectors and be nearby in the embedding space (Figure 3b). Attributes like proximity and direction reflect the similarity of nodes and relationship types. These embeddings are subsequently valuable for link prediction, where predictive ML techniques are used to infer potential links within the KG [62].

One of the foundational approaches for link prediction is matrix factorisation, where the embedding matrices are decomposed into lower-dimensional vectors that capture the latent features of entities and relationships [63]. These latent matrices are subsequently combined to approximate the original matrix and restore original relationships, including missing ones which may be uncovered. Neural-based methods can also identify patterns and assess whether entity embeddings fit together based on their learned features [64]. Link prediction can be particularly beneficial in research because of its potential applications in enriching sparse databases, generating hypotheses, and enhancing question-answering systems [65].

Thus, link prediction methods could serve as powerful gene function prediction approaches, where new links between genes and biological functions could be based on information extracted from scientific literature.

**Synergy of Large Language Models and Knowledge Graphs**

Despite LLMs' exceptional logical reasoning capabilities, they lack the nuanced, domain-specific details [66,67] and can hallucinate, resulting in factual inaccuracy, which warrants caution for users [68,69]. While this can be solved by continual pre-training of LLMs on extensive corpora, such an approach requires vast computational resources and time investment [70]. However, recent advances in computer science show that LLMs and KGs can be used together to minimize hallucinations and use the powerful reasoning capabilities of LLMs for link predictions.

*Taming the flood of literature with LLMs and KGs*

LLMs and KGs can be combined by injecting KG knowledge into LLM prompts or by allowing LLMs to extract information from KGs programmatically [71,72]. Combining LLMs and KGs allows the models to be used for single-hop question answering (QA), where a question can be answered by one edge of the KG (Figure 4a). Multi-hop Question Answering (MHQA) scenarios require interconnected reasoning steps and a nuanced understanding of relationships between various entities within a context to answer more complex questions. Think-on-Graph (ToG) implemented the tandem "LLM⊗KG" paradigm [73], where LLMs search KGs, extracting diverse and multi-hop reasoning paths, where "multi-hop" refers to multi-step connections across different nodes (entities) in KGs. Semantic Graph structures (SG) have proven effective in MHQA [74], where SGs are extracted with edges representing potential reasoning steps, helping the model navigate complex relationships and conduct more accurate reasoning. These and other approaches [75,76] form the basis for LLM reasoning and explain the reasoning process in LLMs, resulting in more precise insights into how conclusions are drawn.

The ability to retrieve factual information and use it to answer complex, multi-hop questions would allow the researchers to extract comprehensive information about gene function rapidly. While such tools are currently unavailable for plant science, examples such as the Educational Question-Answering (QA) systems show an effective integration of cross-data KGs that helps the chatbot generate answers with factual accuracy [77]. We envision that developing a "PlantChatBot" (Figure 4b) would provide truthful, traceable and context-aware answers for gene functions, allowing time-efficient access to the scientific literature.

*Leveraging the reasoning capabilities of LLMs for link prediction*
While initial link prediction methods used vector embeddings, recent research leveraged LLMs' semantic understanding and reasoning capabilities. Chain-of-Thought prompting involves a series of intermediate reasoning steps, and can significantly improve LLMs' reasoning capabilities [78]. LLM possesses an incredible ability for in-context few-shot learning via prompting, where a model can be 'programmed' to a specific task by simply prompting the model with a few input-output exemplars demonstrating the task [79]. LLMs with this approach achieved state-of-the-art accuracy on word problem benchmarks, surpassing even LLMs fine-tuned to a specific task [80]. In the context of gene function prediction and knowledge graphs, an input-output exemplar (Q: Does gene A localize to chloroplast and why? A: Gene A localizes to chloroplast, because chloroplast stroma is a subcompartment of chloroplast) (Figure 4a) could prompt the LLM to predict new links. Indeed, when presented with Figure 4a, GPT-4o predicted: '*Given that Gene B interacts with Gene A, it is reasonable to infer that Gene B might also be localized to the Chloroplast stroma or at least to the Chloroplast. This is because proteins or genes that*

*interact often function in the same or related subcellular compartments to carry out their roles*'. This example demonstrates the potential of using LLMs for link prediction in biological knowledge graphs, where the ability to draw inferences based on complex relationships can uncover novel insights about gene functions and interactions.

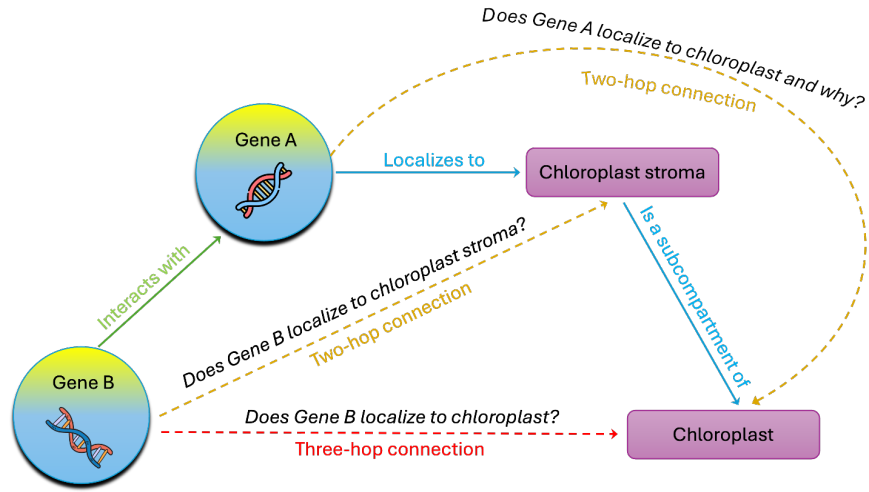

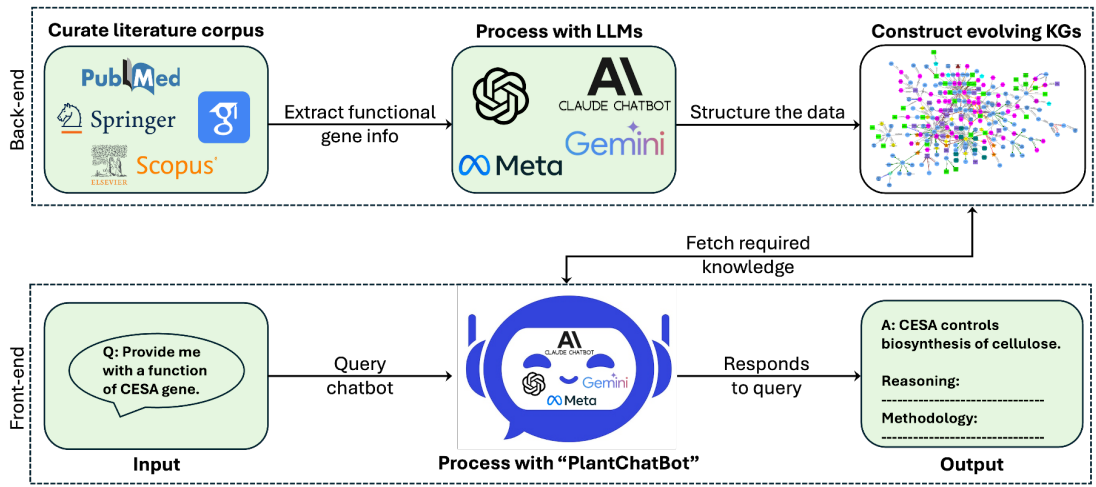

**Figure 4. Link prediction and question-answering system with synergized KGs and LLMs.** (a) The multi-hop link prediction in a KG with LLMs. (b) The pipeline of LLM-based question-answering "PlantChatBot". In the back end, the unstructured knowledge from corpora is analyzed by LLMs to construct evolving KGs, where evolving KG implies continuous updation of KG with new nodes and edges between them as the literature corpus gets updated with new manuscripts. Then, the LLM-based chatbot utilizes the required knowledge from evolving KGs to answer the user's query in the front-end.

## Conclusion

KGs offer a structured database that LLMs can utilize, while LLMs, with their superior logic and reasoning capabilities, can address complex scientific questions posed by researchers. Given the current gaps in our knowledge of plant gene functions, integrating KGs and LLMs would accelerate the excavation of gold standard data from scientific literature, leading to more powerful gene function predictions and testable hypotheses of complex gene/phenotype/environment interactions. Employing tools such as LLMs and KGs to generate gold standard, summarize the burgeoning scientific literature and make predictions would boost gene function annotation and benefit basic and applied research.

## Acknowledgements

MM thanks Singaporean Ministry of Education grant MOE-T2EP30122-0017 'A Kingdom-wide Study of Genes, Pathways, Metabolites, and Organs of Plants' for funding. RSS thanks Singaporean Ministry of Education grant MOE-MOET32022-0002 'From tough pollen to soft matter' for funding. IM thanks IGP-AIX, Interdisciplinary Graduate Programme, Nanyang Technological University, Singapore.